\documentclass{article}

\usepackage{arxiv}

\usepackage[utf8]{inputenc} 
\usepackage[T1]{fontenc}    
\usepackage{hyperref}       
\usepackage{url}            
\usepackage{booktabs}       
\usepackage{amsfonts}       
\usepackage{nicefrac}       
\usepackage{microtype}      
\usepackage{cleveref}       
\usepackage{lipsum}         
\usepackage{graphicx}
\usepackage{natbib}
\usepackage{doi}
\usepackage{graphicx}
\graphicspath{ {./images/} }

\title{Grid cells and their potential application in AI}

\author{ Jason Toy
\\
        somatic \\
	\texttt{jasontoy@gmail.com} \\
}

\renewcommand{\shorttitle}{Grid cells and AI}

\hypersetup{
pdftitle={Grid cells and their potential application in AI},
pdfsubject={q-bio.NC, q-bio.QM},
pdfauthor={Jason Toy},
pdfkeywords={Grid Cells, hippocampus, artificial intelligence},
}

\begin{document}
\maketitle

\begin{abstract}

Since their Nobel Prize winning discovery in 2005, grid cells have been studied extensively by neuroscientists.
Their multi-scale periodic firing rates tiling the environment as the animal moves around has been shown as critical for path integration.
Multiple experiments have shown that grid cells also fire for other representations such as olfactory, attention mechanisms, imagined movement, and concept organization potentially acting as a form of neural recycling and showing the possible brain mechanism for cognitive maps that Tolman envisioned in 1948.
Grid cell integration into artificial neural networks may enable more robust, generalized, and smarter computers.
In this paper we give an overview of grid cell research since their discovery, their role in neuroscience and cognitive science, and possible future directions of artificial intelligence research.
\end{abstract}


\section{Brief History}
The Entorhinal cortex, also called the medial entorhinal cortex (MEC), is an area in the brain that sits between the neocortex and hippocampus in an area called the medial temporal lobe.
In Broadmann's area, it is Broadmann area 28.
Ramon y Cajal described the area in 1902, but at the time thought it was part of the olfactory system used for processing smell information.

In 2005, grid cells were discovered in the entorhinal cortex \cite{hafting2005microstructure}.
Grid cells are a type of neuron that fires at regular intervals as an animal navigates an area.
Grid cells provide a multi-scale periodic representation that functions as a metric for location encoding and is critical for recognizing places for navigation.
When the scientists plotted the points in 2D, they discovered that they formed a grid of tesselating triangles, hence the name grid cells.

The team that discovered grid cells consists of Edvard Moser, May-Britt Moser, and their students.
The team ending up winning a Nobel peace prize in Physiology or Medicine in 2014.
Coincidently, they shared the prize with John O'Keefe who discoverd place cells in the hippocampus.
Together this Hippocampus and Entorhinal cortex (HEC) system are believed to be the main system for building a cognitive map. 
The HEC system is also one of the most frequently studied areas connecting neuroscience with artificial intelligence.

Neuroscienists have been optimistically studying the HEC system trying to unlock how exactly it works to accelerate various fields of science.
Place cells in the hippocampus are special in the brain in that their firing rates act as bookmark for landmarks or an indexing system.
Tthey fire only foor specific locations in an environment.
For example a particular place cell may only fire when at the entrance to work, your favorite parking spot, or at your house.
Place cells are known to fire on 1 to several locations.
Contrast that with grid cells, they retain their basic firing pattern regardless of environment, making them generalized and context independent.  The firing rate forms a hexagonal like grid pattern that tiles to the environment. Grid cells and Place cells feed into each other along with other inputs forming some kind of recurrent network.  

Grid cells have been found in mice, rats, bats, monkeys, and humans. It is believed that they exist in all mammals similar to the neocortex. It is still unknown if grid cells appear in non-mammal animals.

\includegraphics[width=15cm,height=10cm]{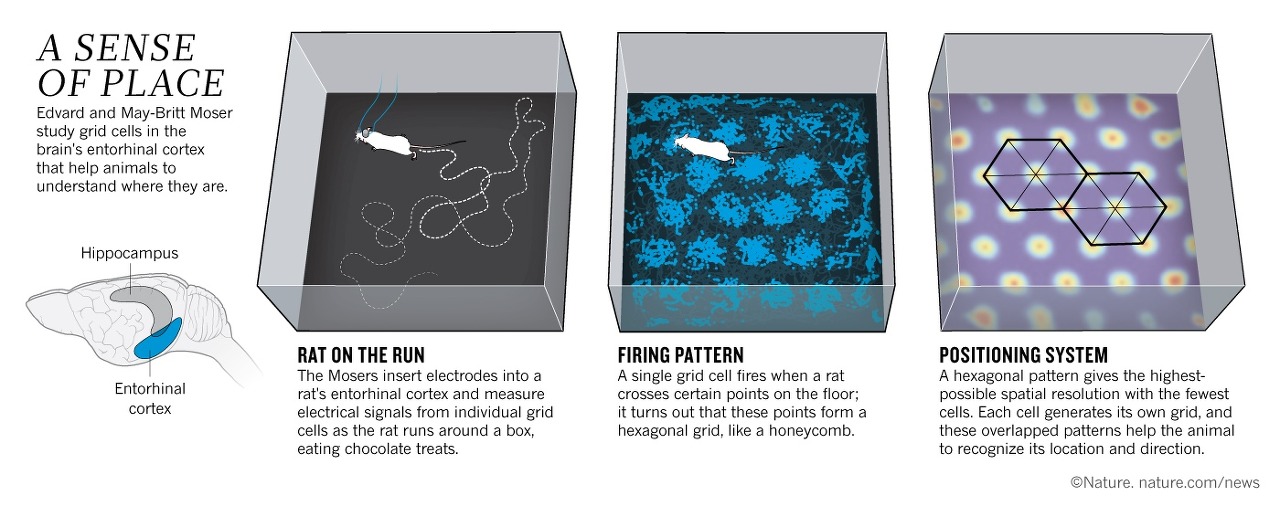}

\section{Human Health}

The EC seems to be a critical part of the mind and brain and its degradation seems to have severe effects on human cognition and health.
In Alzheimer's disease (AD), the EC is one of the first areas of the brain to be noticeably effected, typically with a reduction in volume of the cortex \cite{solstad2008representation}.
AD common symptons are disorientation, easily getting lost, problems with language, behavioral issues, among others. 
Early changes in the EC are used as a strong predictor of AD and these EC changes are commonly used to test for AD \cite{bottero2021key}.
Patient HM (1926-2008) was the most studied patient in neuroscience history \cite{squire2009legacy}.
He had seizures since 10 years old from an accident and at 27 years old had a surgery that removed his hippocampus, entorhinal cortex, and nearby areas.
The procedure stopped his seizures, but it created another problem in that he acquiried anterograde amnesia, he was unable to form new memories.
At the time of his death grid cells were only recently discovered and so there were not many experiments done with him, but the fact that his EC was missing implicates it as being a key system for declaring new memories. 

\section{Grid Cell parameters}

There are 3 parameters that determine a grid cell's firing pattern, the orientation of the grid, the spacing between firing fields (also called the wavelength), and the location (also called spatial phase) of the grid fields. 
Grid cells cluster together into discrete modules sharing orientation and periodicity, but varying randomly in phase.
Currently for all possible configurations of parameters, it is believed that there are under 10 different modules.
Grid cells will maintain their firing patterns even if speed or direction change.   
Grid cells are usually clustered together forming discrete modules of computation.

\section{Coordinates in the brain}
To navigate and interact with our world, our brains must store this information in a way that we can retrieve it and use it in new situations.
One interpretation of the brain is that it is a coordinate translation machine: "a brain is a well-designed machine for the frame conversion to internalize the external world" \cite{arisaka2022grand} and "the egocentric representations of the primary sensory cortical areas must be transformed into an allocentric representation in the hippocampus, and then transformed back to an egocentric motor representation for behavioral output" \cite{byrne2007remembering}.
The brain has been found to store multiple coordinate representations along with different orientations or points of view, mainly allocentric and egocentric orientations.

\includegraphics[width=15cm,height=10cm]{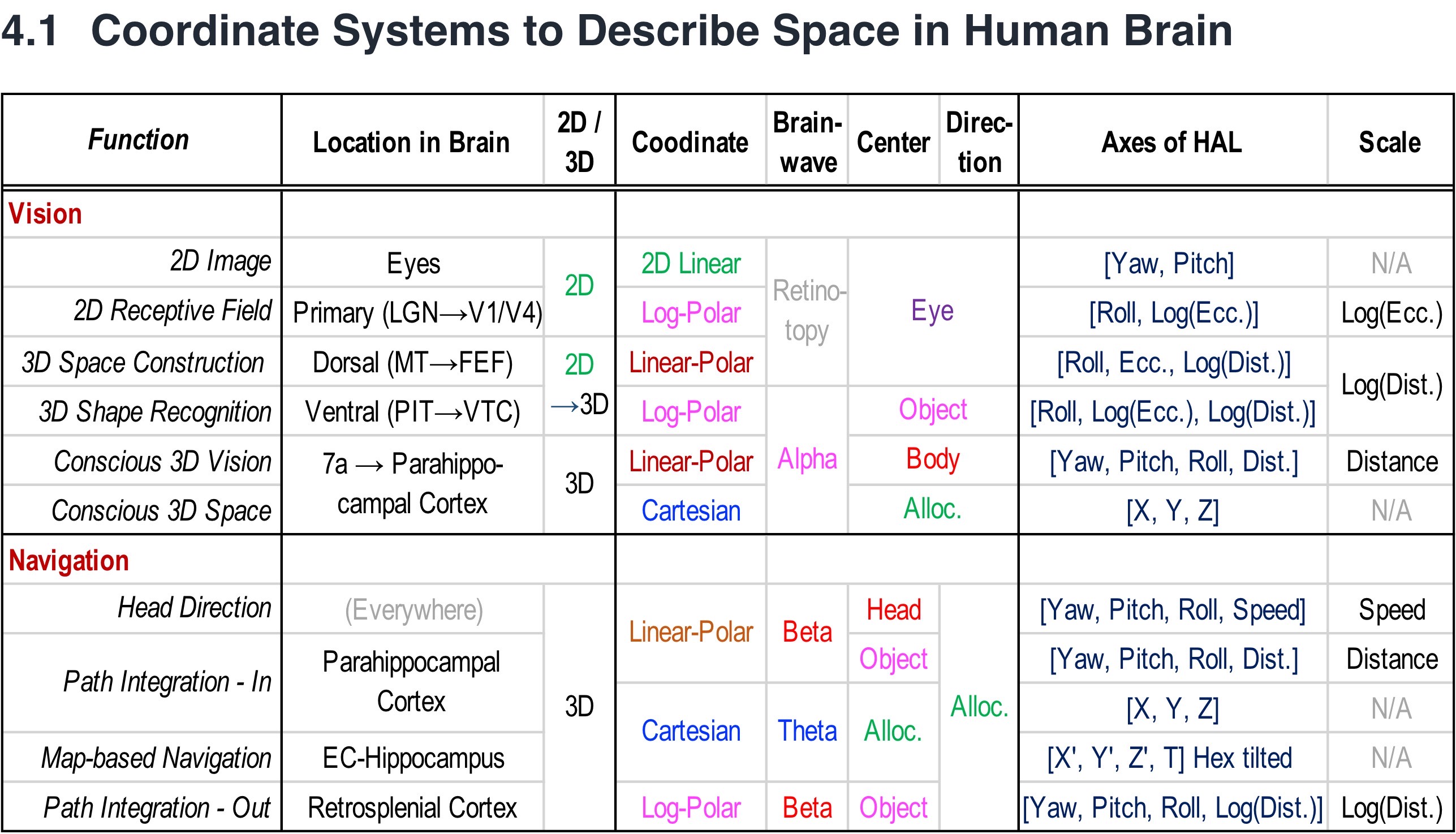}

Grid cells have an allocentric point of view (i.e. world-centered, external world, frame of reference).
Their firing does not encode data from the point of view of the agent or ego.
On the contrary, egocentric neurons fire in relation to the animals point of view such as visual neurons.
For example BVCs are egocentric, they only fire when the agent is near certain borders and within certain angles.
A 3D first person shooter is played from the ego's point of view where as tic-tac-toe is played in an allocentric manner, you can the play game from just knowing the environment coordinates and in any orientation.
In the HEC, it is thought that most of the representations are allocentric, but there are new studies showing that egocentric representations may also be stored. 
In one study \cite{kunz2021neural} found abundant cells in the parahippocampal cortex of human epilepsy patients that they dubbed "anchor cells".
These anchor cells seem to represent egocentric direction towards "anchor points" located in the environment.
 It is thought that these anchor cells are used to transform perceptual signals into allocentric representations.

\section{Other cells in the entorhinal cortex}

The entorhinal cortex is made up primarily of grid cells, but there are other functional cell types as well: head direction cells (HD cells), boundary vector cells (BVCs), and speed cells. There are also cells that form combinations of these cells. Other cell types continue to be discovered.

\subsection{Head direction cells}
The cells fire when the animal's head points in a specific direction similar to a compass. They have a preferred firing direction that they will maintain regardless of the environment that they are in.  HD cells are also found in other regions of the brain such as the thalamus and retrosplenial cortex.  One interesting note is that HD cells continue to fire while an animal is asleep, especially during REM sleep. 
In rats, HD cells appear before the animal has ever opened its eyes.

\subsection{Boundary vector cells}
Boundary vector cells (BVCs), also known as boundary cells or border cells fire when an animal is at specific distances and directions to environmental boundaries such as walls and dropoffs. They are thought to make up a little under 10
The results of their experiments suggest that BVCs are tuned to geometric borders and not by content in the environment.
Evidence has shown that BVCs that landmarks and their borders are critical to maintaining stable grid patterns \cite{hardcastle2015environmental}.
It is thought that BVCs may act as a translation system to grid cells moving to a new map.

\subsection{Speed cells}
Speed cells are neurons who’s firing rate depends on the speed of the organism moving throughout an environment.
Analysis of the MEC has put the population of speed cells to be ~15
They were discovered in 2005.
It doesn’t matter if the organism can see or not, so these cells don’t seem to be influenced by visual cues.
This idea suggests that the speed cells are taking input partially from motor neurons and proprioception neurons to be able to self-generate the speed.
Like grid cells, these cells are also context invariant.
In another experiment, researchers found speed cells were better at predicting future speeds versus past or current speeds, so these cells can be thought of speed prediction cells \cite{kropff2015speed}.
 There have been several studies measuring how far they can measure into the future and studies have shown they predict ~50-80 ms into the future.
It is believed that speed cells are a mix of the other cell types in the MEC.
Previous experiments suggest that speed cells are used to dynamically update grid cell activity and enable efficient path integration acting as error correct codes.

On top of all these distinct cell types found in the MEC, there are small populations of cells that work as conjuctives of the main cell types, meaning they function as both grid cells and speed cells or grid cells and head direction cells.

\section{Grid cells and place cells}

Grid cells mainly feed into place cells, but its not that simple. They both feed into eachother forming a recurrent network.
Experiments have shown that in rats place cells develop before grid cells \cite{langston2010development}.
Several studies have shown MEC neurons receive input from hippocampal place cells: Place cells in the hippocampus emerge earlier during development than grid cells in the MEC, grid cells lose their tuning pattern when the hippocampus is deactivated, and both the firing fields of place cells and the spacing and field size of grid cells increase along the dorso-ventral axis.
Moreover, entorhinal stellate cells, which often exhibit grid-like firing patterns, receive a large fraction of their input from the hippocampal CA2 region, where many cells are tuned to the location of the animal.
In one paper, researchers propose that place cells feed into grid cells to act as a lower dimensional reprsentation of place cells in a principle component analysis (PCA) form \cite{dordek2016extracting}.
Some experimental evidence has shown that place cell input to grid cells acts as a correcting mechanism for drift and anchors grids to environmental cues  \cite{barry2007experience}.
Another interpretation is that grid cells may be understood as the eigenvectors of the relationship between place cells.
Another interpretation of the interplay between grid cells and place cells is that grid cells act as a coding schema or lookup table for accessing specific memories stored inside of place cells.
In another paper \cite{bush2014grid} ran experiments that show that grid cells to place cells firing is not a successive processing pipeline, but instead that place cells are taking multiple inputs from different areas and that grid cells  provide a complimentary and interacting signal to support reliable coding of large-scale space.
Our understanding of the interaction of grid cells and place cells is still primitative and much more experiments need to happen.

\section{Other Representations}

Spatial navigation is thought to be the main focus of grid cells and the entorhinal cortex, but with its generalized grid code that fires consistently regardless of the environment, many have wondered if this grid code is used to represent other neural representations.
 In the following paragraphs we will describe experiments that focus on grid codes that represent other sensory signals and representations.
The results of these experiments seem to show that grid cells may have evolved for other uses acting as a form of neural recycling.

\subsection{Olfactory}

In another paper \cite{bao2019grid}, the authors did experiments to see if grid cell fired for navigating in a 2-D odor space. The main experiment consisted of having patients imagine moving through a 2D space where each location has a different odor all while being in an fMRI machine. They note: “In this study, we tested the hypothesis that human subjects, using only the sense of smell, could navigate through a two-dimensional olfactory space. When provided with a start odor location and route (trajectory) instructions, subjects were able to imagine and predict their perceptual translocation in this odor space. Odor navigation was associated with hexagonal grid-like coding in vmPFC, APC, and ERC, with behavioral performance scaling with the robustness of entorhinal responses across subjects. These findings mirror the behavior relevance of grid-like units in navigation of physical and abstract spaces (Constantinescu et al., 2016; Doeller et al., 2010; Kunz et al., 2015) and highlight the idea that the human brain has access to internalized representations of odor mixture arrays to guide spatial orientation and route planning.” The findings are encouraging, but not strong due to experiment design and results.

\subsection{Auditory}
There have been multiple experiments trying to understand if and how grid cells also represent auditory sensory information.
In a paper by \cite{aronov2017mapping} they wanted to know if grid cells would respond to auditory stimulation and produce the same hexagonal firing patterns typically seen when navigating an environment.
Previous multiple experiments have shown that hippocampus place cells respond to distinct auditory sounds.
In their experiment, they trained a rat to receive rewards by controlling a joystiq to play sound frequencies and when a range of a certain sound frequencies was reached and then released, rewards would come out.
The rat was able to successfully control the joystiq and receive numerous rewards.
The experiment results saw that some of the grid cells fired because of the behavioural tasks and so to test this more, they tried passive playback to see if that changed the firing. Place cells wouldn't fire, I'm not sure about grid cells.

In a paper titled "Are Grid-Like Representations a Component of All Perception and Cognition?", Chen et al. analyze previous grid cell experiments and argue that grid responses may be universal in the brain.
Based off the facts that multiple mammals use echolocation for navigation and conjuctive sound and phyiscal feature neurons have been found, they predict that more grid-like firing patterns will be discovered in the auditory cortex \cite{chen2022grid}.

\subsection{Overt attention}

In another study \cite{killian2012map}, the team of researchers did an experiment on monkeys showing that grid cell firing patterns appeared when a monkey analyzes a scene without locomotion.
The monkeys had their head fixed, but were allowed to shift their eyes.
The monkeys were then shown various pictures while the researchers recorded the eye scan paths and the firing rates.
The researchers then processed the data calculating spatial autocorrelegrams and gridness scores.
Their results suggest that grid cells may also represent visual exploration without requiring an actual visit to that particular location.

In another study, a different group of researchers built off \cite{killian2012map}, but instead used humans an fMRI data \cite{julian2017evidence}.
Participants were tasked with looking at images of letters to try and find a target letter.
Their results also showed that humans exhibited grid patterns in the MEC while doing the visual search task.

\subsection{Covert attention}

In a study from Wilming et al. titled "Entorhinal cortex receptive fields are modulated by spatial attention, even without movement", researchers showed that movement of covert attention, without any physical movement elicited grid cell responses \cite{wilming2018entorhinal}. 
The researchers trained rhesus macaque monkeys to maintain a fixed gaze on the center of a monitor while a small dot moved around.
The monkeys had to press a button when the dot changed colors.
The researchers took neuronal firing recordings from both the MEC and hippocampus and analayzed the data.
They found no grid cell firing activity in the hippocampus while finding grid cell firings in the MEC.
Their results show strong evidence that physical movement through an environment is not required to elicitt grid cell activations and that the grid cells can represent the location of attention.

\includegraphics[width=16cm,height=10cm]{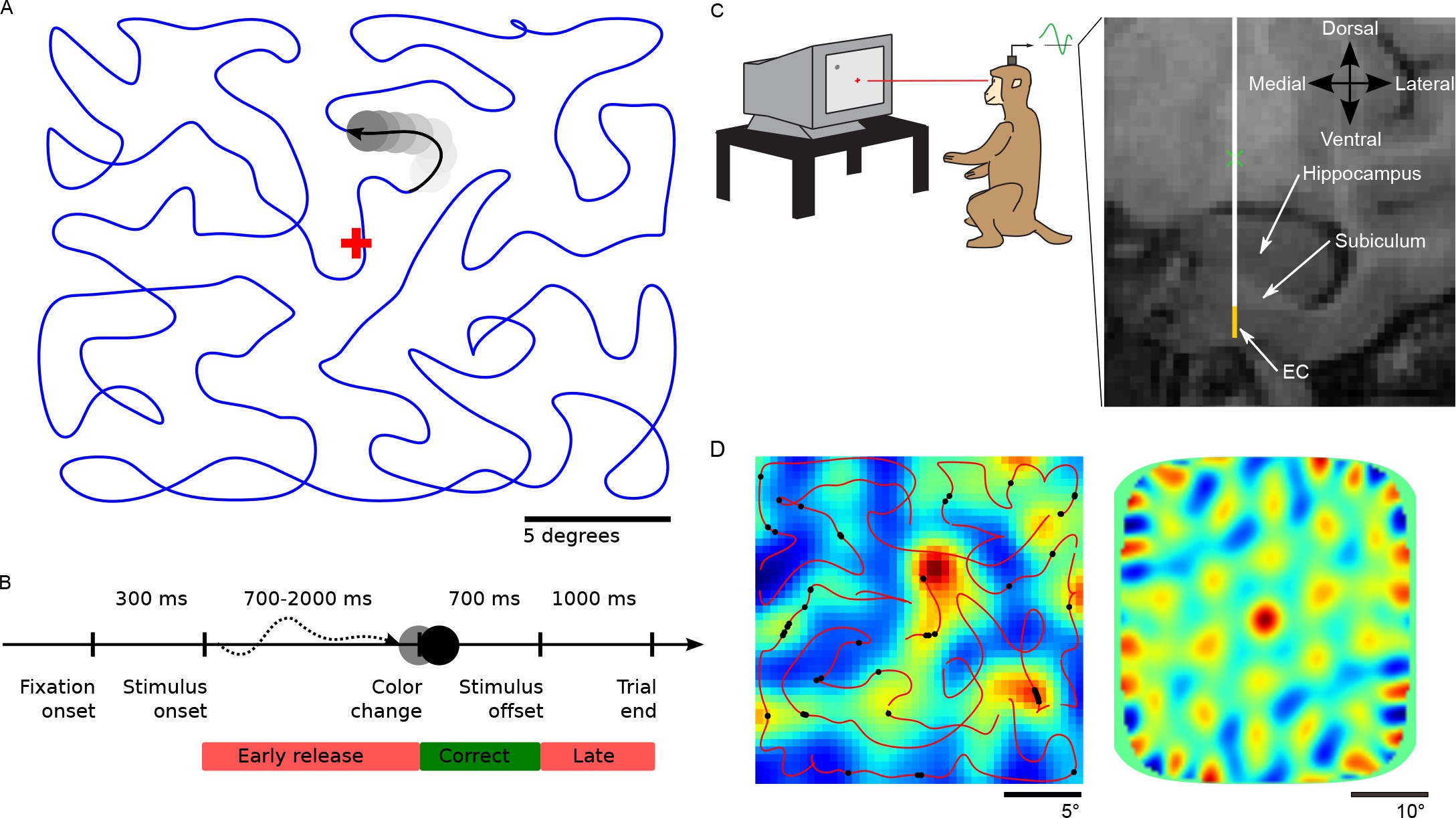}

\subsection{Mental simulation}
In an experiment from \cite{bellmund2016grid}, the team tested humans using fMRI to see if imagining movement would generate a grid like code.
Participants were allowed to explore a large-scale realistic virtual reality city for 10 minutes.
Then they were put into an fMRI machine and asked to imagine navigating between different landmarks without access to the city.
Their analysis of the data did find grid like codes in the MEC.
The authors believe their findings suggest that grid cell systems are used for goal-directed decision making and that spatial computations may be fundamental to many forms of mental computation.

\includegraphics[width=16cm,height=10cm]{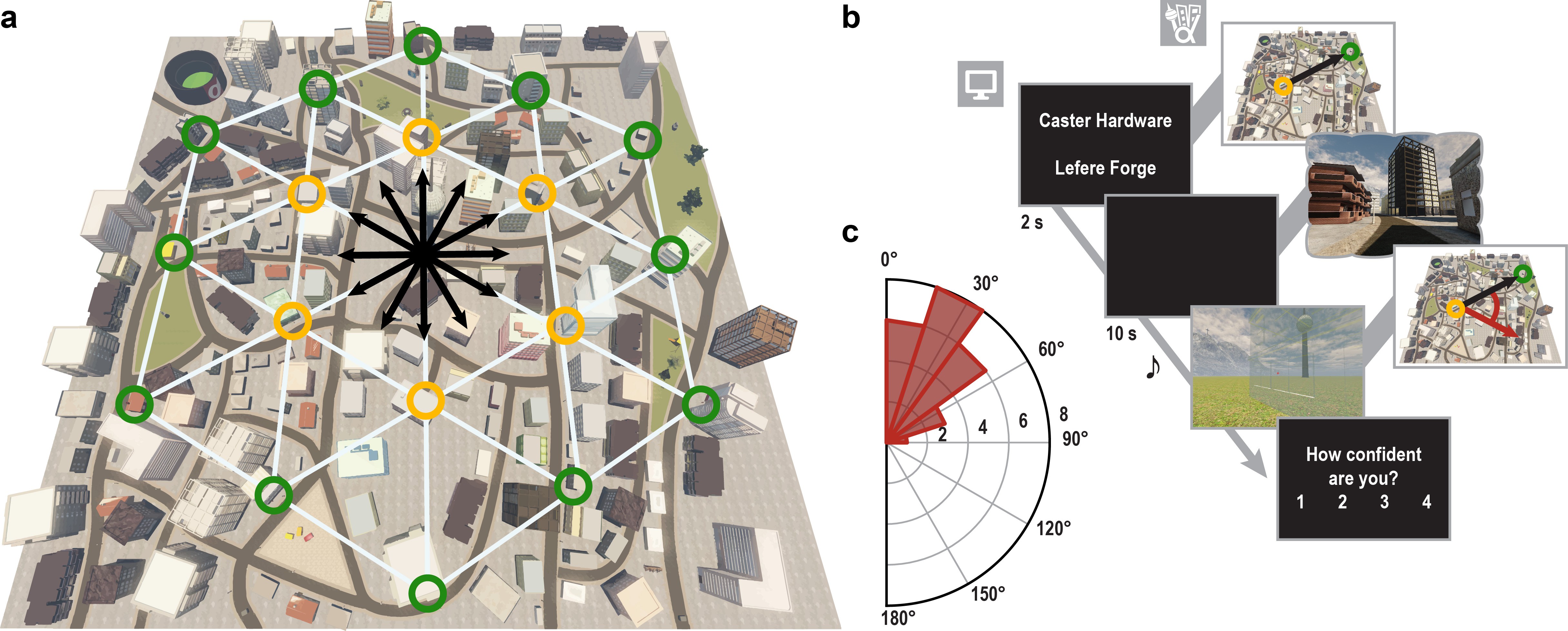}

\subsection{Abstract concept space}
In a paper by Constantinescu et al., the researchers wanted to know if the grid codes from grid cells can be used to organize non-spatial conceptual representations \cite{constantinescu2016organizing}.
The researchers contrived an experiment where human patients where trained to associate various lengths of bird necks and bird feet to different christmas symbols.
These various bird attributes combined to create a "continous 2D bird space" that the patients were not consciously aware of.
After patients were trained to walk the whole bird space over several sessions, patients were put into an fMRI machine as they watched videos of birds morphing from predefined neck:leg ratios. 
The patients were then instructed to imagine to choose which christmas symbol would be chosen if the bird continued to morph with the same neck:leg ratios.
From the fMRI data, they were able to see hexagonal symmetry and grid like firing patterns in the MEC.
An important note is that they measured this grid code in the neocortex as well, specifically the medial prefontal cortex (mPFC).
Their results show that humans grid cells are firing with their typical hexagonal grid code during abstract non-spatial cognition suggesting that grid cells.

There are more experiments we would like to see.
Would we get similar results if we were to wire electrodes directly to the MEC of humans, such as epileptic patients.
This would give us more unequivocal evidence.
Could we construct a similar 2D cognitive space that a rat could understand and then wire electrodes directly to their grid cells.
Could we do a similar experiment, but with a 3D space and analyze the results.

To date, direct evidence of single-unit grid representations in conceptual spaces has not yet been discovered, partially due to the limited accessibility of human brains in clinical settings.

\subsection{Semantic word meaning space}

In another paper by \cite{vigano2021grid}, a group of researchers wanted to see if a word space with multiple dimensions would elicit grid codes.
The team constructed 9 words with no meaning such as  pad, tid, gal, vig, jot and attached to them different sizes and sound pitches to associae audio-visual objects with the words.
31 patients were then trained in several different tasks to differentiate those words properties without ever knowing that they were mapped to a 2D grid.
Then the patients were given a comparison task where two words were sequentially shown and then asked one of two questions: “has there been an increase, decrease, or no change in size?” or “has there been an increase, decrease, or no change in pitch?”
While they answered those questions, their brains were scanned in an fMRI machine.
Participants got the answer right 89
The fMRI results show that grid cells did fire in their typical hexagonal pattern.
The team believes that their results show that grid cells can process purely symbolic stimuli.

\includegraphics[width=16cm,height=10cm]{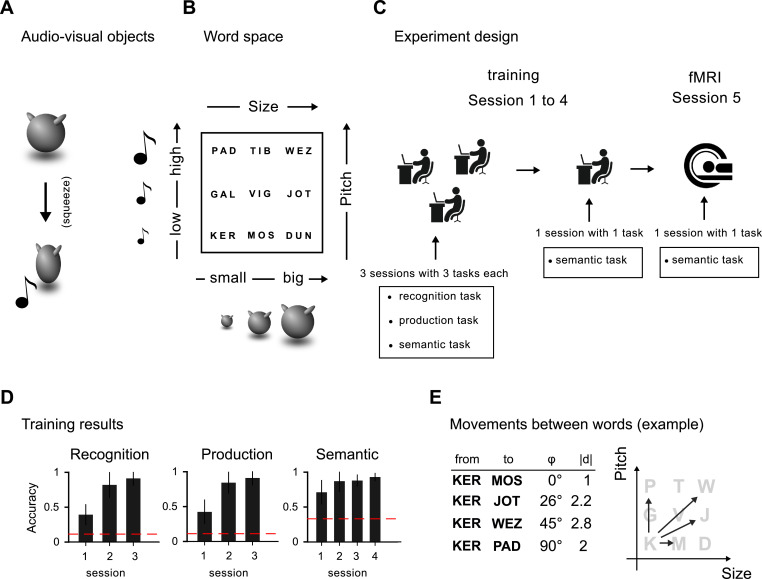}

\subsection{Time}

To go from one location to another requires movement which must happen over time.
Time and space are deeply intertwined, for example in physics, spacetime is a mathematical model that merges the three dimenions of space with one dimension of time into a single 4D manifold.
Scientists have wondered if grid cells also represent time.
It has been known that cells in the hippocampus can represent abstract time directions \cite{eichenbaum2014time}.
In a study from Kraus et al., the group of researchers studied rats in a treadmill running at various speeds and distances.
They built several models of the data to factor out location and found that the grid cells were actually weakly association with location.
They found that most grid cells where either strongly correlated with a combination of time and distance or either just time or distance.
Their findings suggest that in the absence of external queues, grid cells integrate self-generated distance and time information and that grid cells with the hippocampus represent both temporal and spatial dimensions.

\subsection{3D Representations}
The majority of grid cells studies has been on 2D physical spaces.
All animals are moving in a 3D space, especially animals that are fly or live in the sea such as bats and dolphins.
So the next natural step has been to understand grid cells in 3D. 
Researchers expected the same periodic firing pattern in a lattice like structure to appear while traversing 3D space.
Place cells and head direction cells have already been shown to directly generalize to 3D.
Many experiments have been run to understand 3D grid codes, but while 2D representations of grid code is undeniable, finding 3D representations has remained an elusive problem.  
In a paper from \cite{ginosar2021locally}. they recorded MEC grid cells of flying bats to understand 3D grid code patterns.
The team created an enclosure of size 5.8m x 4.6m x 2.7m with several feeding areas placed throughout the enclosure to encourage the bats to fly to different locations.
They expected the hexagonal grid code to move uniformally into 3D like a stack of oranges in a grocery store representing optimal packing, but the results show that in 3D there was no neat uniform pattern. 
 They claim their results show that in 3D, many cells have multiple 3D fields and that in 3D the grid code has local structure and no global lattice, but when constrained to 2D will form a hexagonal global lattice.
Given the findings, they suggest that global spatial periodicity is not central to grid cells.
In another paper \cite{grieves2021irregular}, researchers tested 3D grid cells with rats exploring a 3D space.
Their results are similar and found that the grid fields were irregulary arranged, sparser, large, and more variably sized.
They suggest that grid codes may not need to be tiling or tightly packed to support spatial computations.
Therefore new models need to be created to describe grid cell hippocampus interactions.
There are several computational and mathematical models that have been developed that do fit to grid cell firing patterns in both 2D and 3D.
If we are to get to a multidimensional abstract conceptual space using grid cells, then more 3D and multidimensional experiments with time will need to be performed.

\includegraphics[width=16cm,height=10cm]{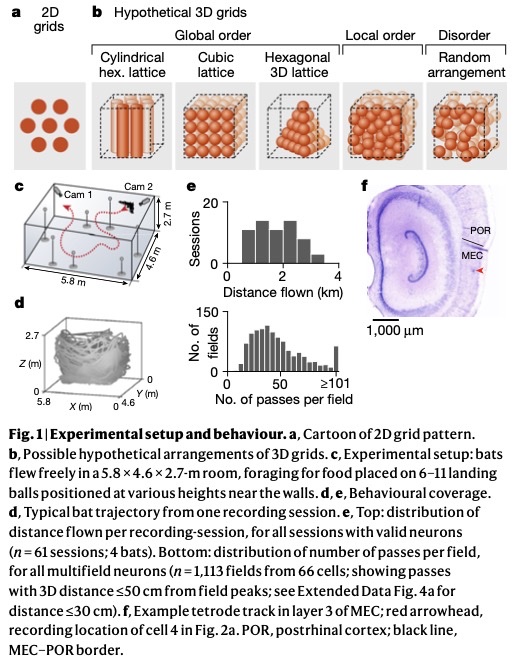}

\section{Grid codes in other areas of the brain}

There have been multiple experiments that show grid codes similar to grid cells exist in other parts of the brain such as the neocortex.
In the paper "Evidence for grid cells in a human memory network", They wanted to see if they could find grid-like firing using fMRI signal, which reflects changes in metabolic activity across large portions of the brain resulting in very coarse readings of neuronal activity. They designed an experiment using a virtual environment in a way that they could measure 3 factors from the fMRI data. 1) The angular orientation relative to the environment, 2) The running direction, and 3) The running speed. They believe they did find grid-like cell activity in multiple neocortex regions such as the visual cortex and retrosplenial cortex. They note “Because we ca only measure effects of direction and speed (not location) in fMRI signal, our findings could reflect the presence of grid cells, or movement-related responses from head direction, or ‘conjuctive’ directional grid cell…"
In another paper, "Direct recordings of grid-like neuronal activity in human spatial navigation", researchers surgically implanted electrodes into epilepsy patients.
They had recordings for 893 individual cells in the hippocampus, amygdala, parahippocampal gyrus, EC, and cingulate cortex. They had those patients perform navigation tasks in a virtual environment cut up into a 28 x 28 array. They came up with a gridness score to classify if the neurons had grid-like codes.
Their results found that of all the recorded cells, 14

In a paper by Constantinescu et al, They designed an experiment for humans to navigate through an abstract conceptual presentation consisting of “bird space”, arrangements of different lengths of bird necks and legs. 
They then measure the participants over several sessions spanning weeks using fMRI.
 In the fMRI signal, they found hexagonally symmetric activity in the vmPFC.

In another paper by Long et al, scientists measures 2025 cells from 8 implanted rats that were able to freely move.
They did 287 recording sessions of their neuron firing activity.
They found that cells known primarily to be in the entorhinal cortex such as head direction cells and border vector cells, were found in the somatosensory cortex.
Of all their cell recordings, they found 3.55
They found that these grid cells were a little nosier and less prevalent than the other cell types they found.
They hypothesize that this may be due to other factors such as proprioception.

With multiple experiments showing grid codes in other parts of the brain, it seems like grid like navigation code has multiple uses for intelligence.
Hawkins et al. believe that the neocortex uses grid cells as a fundamental part of human level intelligence and are trying to build new computer algorithms that use neocortical grid cells \cite{hawkins2019framework}.

\includegraphics[width=16cm,height=10cm]{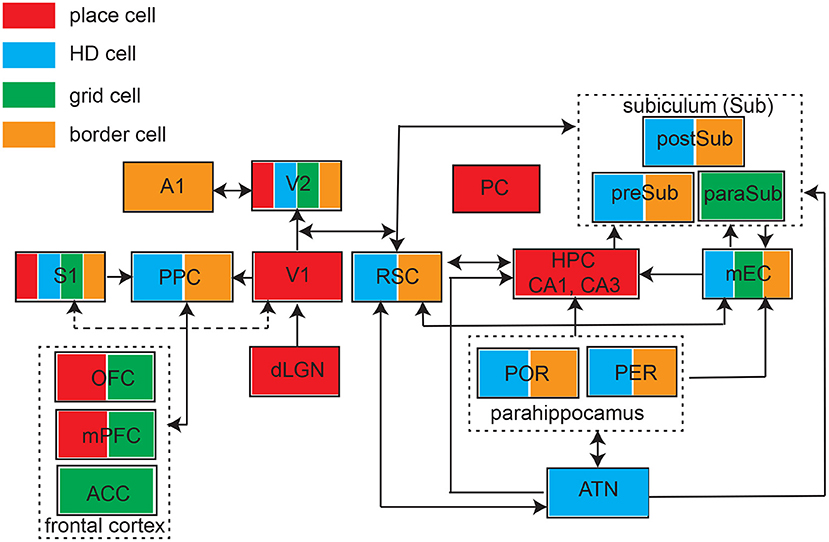}

Scientists have discovered MEC neurons such as grid cells in other regions of the brain.
It is thought that grid cells may appear in more regions of the brain, but it is difficult to detect when not explicitly looking for grid-like responses. 
Specific grid cell experiments must be constructured when looking for grid responses.
Moreover, Many neurons in the neocortex and somatosensory regions seem to respond to more diverse stimulus making it hard to analyze.

\section{Tooling}

The majority of grid cell experiments done on humans are done with functional Magnetic Resonance Imaging (fMRI).
fMRI is used most often to measure blood-oxygen-level-dependent (BOLD) signals, or in other words changes of oxygen levels in the blood as a proxy for neuron activity. 
It shows where blood is being sent in the brain, presumably because neurons in that area are more active during a mental task.
It cannot directly measure neuron communication through electric or chemical impulses.
It is noninvasive which means its can be used on practically any human.
From a technical point of view, it has multiple problems. 
The fMRI acquisition time is meausuring blood flow response 6 seconds after neural activity occurs.
This is eons compared to the speed of computation that neurons communicate at where as electrical activity measurements are in real time .
To give an analogy, its like measuring how a boat engine functions by measuring distant waves the boat produces as it moves 6 seconds later.
fMRI resolution is measured in voxels which is a 3d pixel that ranges anywhere from 1 to 5 cubic millimeters in size.
A voxel contains approximately 1 million neurons.
When fMRI measurements are occuring, the patient must remain completely still, otherwise it will throw off the measurements.
After the raw data is acquired, mathematical analysis must occur to interpret the data which can lead to other types of statistical problems.
In a real, but satirical study titled "Neural Correlates of Interspecies Perspective Taking in the Post-Mortem Atlantic Salmon: An Argument For Proper Multiple Comparisons Correction", a dead salmon was placed in an fMRI machine and then showed pictures of humans in different emotional states.
The authors were able to show the dead salmon had significant increases in "activation" highlighting the importance of correcting your statistics in fMRI studies \cite{bennett2009neural}.
Only when there are patients with severe problems such as epileptic patients can direct measurement of neuronal activity occur in humans.
So a large poriton of grid cell experiments continue to happen on rodents and other mammals.
Of note, these measurement problems apply to the majority of human neuroscience experiments, not just grid cell experiments.
If we are to improve our undertanding of grid cells and the brain, much more precise and improved tools will need to be invented to examine neuronal communication.

\section{Analysis of firing patterns}
In the paper that discovered grid cells from Hafting et al., the team used a technique called spatial autocorrelelograms (SAC) to analyze the neurons' firing rates and discover the heaxagonal grid code.
SAC is a spatial measurement of similarity (correlation) between nearby observations.
The team implanted electrodes into the MECs of rats and recorded them while foraged in an open environment.
They cut the environment into square bins and then recorded when the grid cell fired (generated an action potential) and added 1 to the counter for that bin creating a histogram of the firing rates over locations.
Then for each cell's rate map, a copy of the same map is applied on top and compared.
The copy of the rate map is shifted to the next peak for every coordinate on the 2D map and the autocorrelation value is calculated.
By calculating these spatial autocorrelelograms, it becomes clear that the cell response is represented by a grid.
SAC is the most common method for analyzing grid responses in neuronal firing data, but it sometimes generates false positives.

\section{Artificial Intelligence}

While deep neural networks have made great progress in image processing, classification, audio generation, protein folding, etc, there are still obvious gaps where they completely fail.
ANNs are still not able to generalize and with their large data requirements, have been shown to be mainly doing interpolation and not extrapolation.
If movement is a defining characteristic of intelligence: hunting, evading, navigating home, finding a mate, etc and grid cells are a key component of path integration and movement, then there may be computational properties that we want to emulate in artificial neural networks (ANNs).
Neuroscience research has shown us that grid cells and their firing patterns can represent both spatial and non-spatial sensory representations.
Most of the work with grid cells has happened in the neurosciences, but there has been some early work trying to incorporate grid cells with ANNs.
Research has shown that grid cells coding range is very large and that their spatially period response is combinatorial in capacity \cite{fiete2008grid}.
Some theories believe that this code is used as lookup table to access specific memories inside of the hippocampus.
Other research has shown that the grid code acts as a form of error correction while navigating a space.

In a paper from \cite{banino2018vector}, researchers tested out path integration using a RNN that had cells that fired and looked just liked grid cells.
The authors built a deep learning LSTM model trained on locations, velocity, and head direction to predict where the agent would move to.
They used spatial autocorrelograms to analyze the firing rates of the artificial neurons and found that the LSTM produced cells that behaved just like hexagonal grid cells and BVCs.
They then took the LSTM model and used it as a base layer for a A3C reinforcement learning agent that navigated an environment. 
They found that the RL agent was able to navigate and take previous unseens paths faster than all the models they compared it to.
Of note, they were only able to produce grid like firing patterns when they incorporated dropout.
In our own experimentation, we were only able to get grid patterns when we trained at the same batch size of 10 that they did, anything else higher drastically reduced the amount of grid cells appearing.
The findings suggest that grid cells are a natural computational unit that will appear naturally in navigational problems. Deepmind has subsequently went on to patent this computational pattern in "Performing navigation tasks using grid codes" \cite{banino2020performing}.

In another paper that builds off the work of \cite{banino2018vector}, researchers wanted to see what was the minimal neural network architecture needed to make grid like cells appear.
They tested several different architectures ranging from a 1 layer NN, LSTM, to RNN and found that all of those models were able to produce grid like patterns IF they followed certain constraints.
They found the minimal needed architecture is "a simple linear, place cell encoding objective" with non-negative constraints.
They also found that the shape of the grid pattern depends specifically on the shape of the cell tunng curve.
They found their findings confusing seeing that several previous paper needed much more to make grid cells appear <\cite{sorscher2019unified}.

Another group of researchers ran tests on hundreds of different ANNs optimizing for path finding and found that most networks did not generate grid cells and that specific hyperparameters along with using an optimization method using difference of Gaussians is what would make grid cells appear contradicting previous studies \cite{schaeffer2022no}.
In another paper, a team of researchers implemented grid cells in an ANN to study how a network may process arbitrary sequence of input samples similar to how bilogical eyes use saccades \cite{leadholm2021grid}.
They make the case that ANNs can only handle sequential processing while humans handle out of sequence recognition easily and naturally.
They explored grid cell path finding mechanisms for their model and found that it performed similarly to a CNN when tested on the MNIST dataset.

The Transformer architect is currently one of the fastest growing innovations in ANNs and have made many breakthroughs in language modeling, image processing, and controlling computers.
In a paper from Whittington et al., they have shown that when a small modification is made to the transformer architect, they learn and act like grid cells and place cells \cite{whittington2021relating}.
They modify the transformer architect by adding recurrent positional encodings which creates a form of path integration.
The core mechanisms of transformers is selt-attention where each element is trained to attend to all the other elements and update itself accordingly.
Their findings help make a stronger link between neuroscience and machine learning and a potential roadmap for further exploration in machine learning grid cell integration.

\includegraphics[width=16cm,height=10cm]{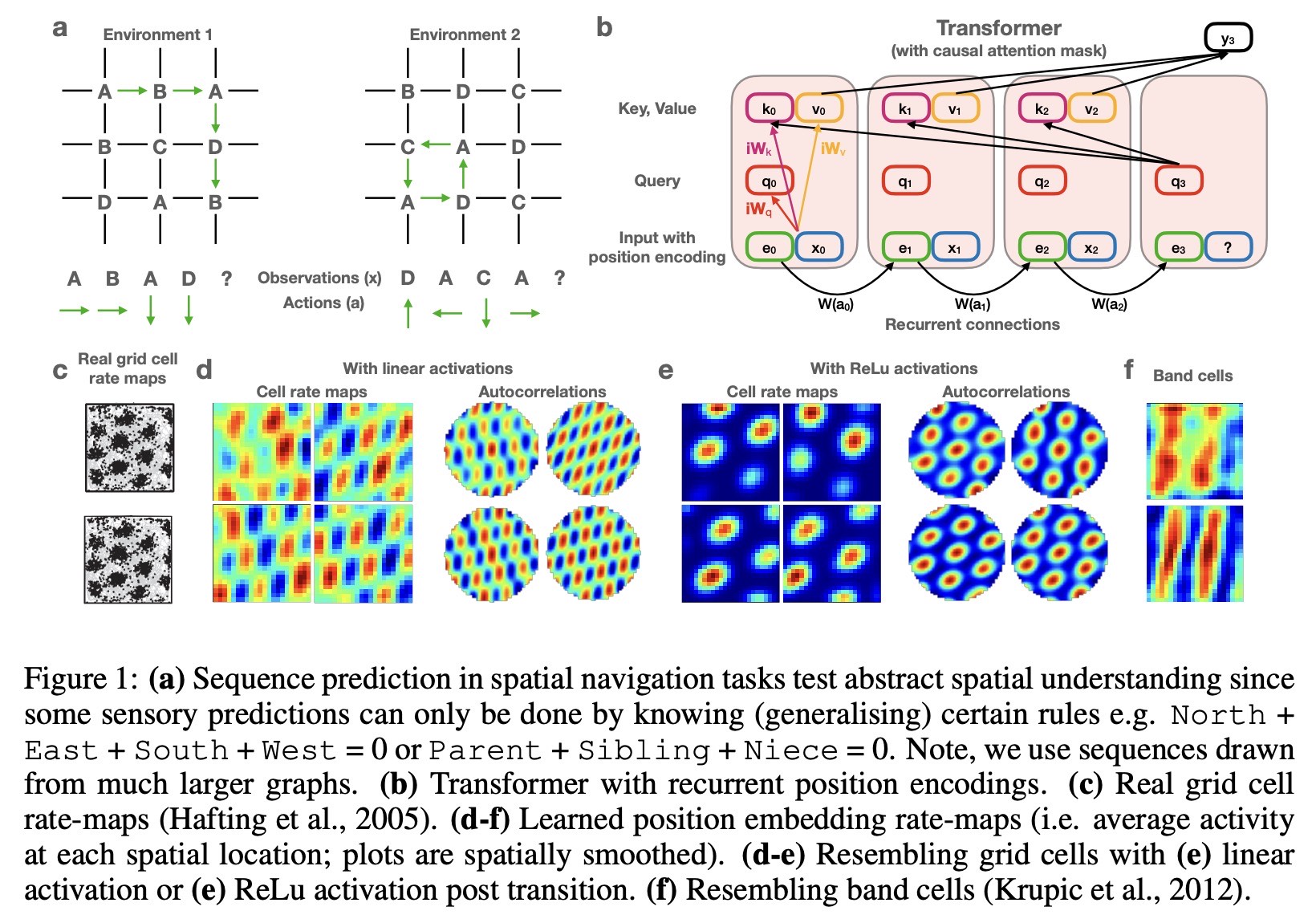}

To incorporate grid cells into ANNs in a meaningful way, the computational and mathematical properties must be understood.

\section{Discussion}
We hope that this overview of grid cells helps you understand their importance to neuroscience, cognitive science, and artificial intelligence.
For neuroscience it gives researchers a window deep into the cognitive processing pipeline that allows us to experiment and measure directly inputs to outputs.
In a paper from \cite{joan2017could} titled "Could a Neuroscientist Understand a Microprocessor?", they argue that the complexity of the brain and the crude tools to run experiments often produce results that are not meaningful towards the understanding of neural processing systems. 
They show this by using neuroscience techniques to analyze a human engineered microprocessor and the results are that they could not meaningfully understand its information processing systems.
With grid cells, we have a method to measure directly part of the information processing pipeline giving us a direct mapping of movement inputs to grid code outputs.
For cognitive science, its generic code along with all the other MEC cells show that the brain does have basis cells for computation that lead to other computations in the hippocampus and other places. 

In artificial intelligence research, artificial neural networks (ANNs) have made huge progress in the past decade with summation point neurons as the fundamental building block.
Some question whether the trend of "just scaling up" by adding more computation and parameters will lead to more intelligent models.
A neuron on average transmits action potential or operations at the rate of 1000/sec while a CPU can do 10 billion/sec.
The method and efficiency of the way human brains process information versus how computers process information is fundamentally different.
The past few decades of AI research has shown that pure computer science based AI research has led to many dead end paths such as expert systems, ontologies, symbolic AI, among others. 
If we are trying to emulate human level intelligence, then studying human brains seems very reasonable.
Demis Hassabis has been advocating this approach of neuroscience-inspired AI to reach the next level of intelligent machines and they have taken this approach at DeepMind \cite{hassabis2017neuroscience}.
Many innovations in deep learning were inspired from neuroscience findings such as convolutionl neural networks (CNNs), point summation neurons, and attention mechanisms.
Grid cells offer the potential to be another building block to build new types of ANNs in a wide variety of applications.

Its abstract hexagonal firing code continues to be studied to understand its relation to other cognitive processes such as language comprehension and concept formation as a form of neural recycling. 
With grid cell experiments showing they can fire for organizing concepts into a conceptual space and experiments showing grid codes appearing in multiple places in the brain, there are many possible experiments that could be explored integrating grid cells into ANNs.
In the paper from \cite{leadholm2021grid}, they showed an example of using grid cells to implement non sequential image processing for classification, this is a promising direction as current CNNs must scan a whole image before a confident classification appears making computation expensive and slow. 
Compositionality, the ability to understand complex meaning from basic units, has been extensively studied in AI for decades, but has made little progress.
In the future, we would like to see experiments incorporating grid cells into word2vec or other semantic models to enable compositionality.
For researchers trying to advance the state of the art in ANNs, grid cell firing codes are a prime target to investigate.

\bibliographystyle{unsrtnat}


\begin{thebibliography}{35}
\providecommand{\natexlab}[1]{#1}
\providecommand{\url}[1]{\texttt{#1}}
\expandafter\ifx\csname urlstyle\endcsname\relax
  \providecommand{\doi}[1]{doi: #1}\else
  \providecommand{\doi}{doi: \begingroup \urlstyle{rm}\Url}\fi

\bibitem[Hafting et~al.(2005)Hafting, Fyhn, Molden, Moser, and
  Moser]{hafting2005microstructure}
Torkel Hafting, Marianne Fyhn, Sturla Molden, May-Britt Moser, and Edvard~I
  Moser.
\newblock Microstructure of a spatial map in the entorhinal cortex.
\newblock \emph{Nature}, 436\penalty0 (7052):\penalty0 801--806, 2005.

\bibitem[Solstad et~al.(2008)Solstad, Boccara, Kropff, Moser, and
  Moser]{solstad2008representation}
Trygve Solstad, Charlotte~N Boccara, Emilio Kropff, May-Britt Moser, and
  Edvard~I Moser.
\newblock Representation of geometric borders in the entorhinal cortex.
\newblock \emph{Science}, 322\penalty0 (5909):\penalty0 1865--1868, 2008.

\bibitem[Bottero et~al.(2021)Bottero, Powers, Yalamanchi, Quinn, and
  Potashkin]{bottero2021key}
Virginie Bottero, Dallen Powers, Ashna Yalamanchi, James~P Quinn, and Judith~A
  Potashkin.
\newblock Key disease mechanisms linked to alzheimer’s disease in the
  entorhinal cortex.
\newblock \emph{International journal of molecular sciences}, 22\penalty0
  (8):\penalty0 3915, 2021.

\bibitem[Squire(2009)]{squire2009legacy}
Larry~R Squire.
\newblock The legacy of patient hm for neuroscience.
\newblock \emph{Neuron}, 61\penalty0 (1):\penalty0 6--9, 2009.

\bibitem[Arisaka(2022)]{arisaka2022grand}
Katsushi Arisaka.
\newblock Grand unified theory of mind and brain-part i: Space-time approach to
  dynamic connectomes of c. elegans and human brains by mepmos.
\newblock 2022.

\bibitem[Byrne et~al.(2007)Byrne, Becker, and Burgess]{byrne2007remembering}
Patrick Byrne, Suzanna Becker, and Neil Burgess.
\newblock Remembering the past and imagining the future: a neural model of
  spatial memory and imagery.
\newblock \emph{Psychological review}, 114\penalty0 (2):\penalty0 340, 2007.

\bibitem[Kunz et~al.(2021)Kunz, Brandt, Reinacher, Staresina, Reifenstein,
  Weidemann, Herweg, Patel, Tsitsiklis, Kempter, et~al.]{kunz2021neural}
Lukas Kunz, Armin Brandt, Peter~C Reinacher, Bernhard~P Staresina, Eric~T
  Reifenstein, Christoph~T Weidemann, Nora~A Herweg, Ansh Patel, Melina
  Tsitsiklis, Richard Kempter, et~al.
\newblock A neural code for egocentric spatial maps in the human medial
  temporal lobe.
\newblock \emph{Neuron}, 109\penalty0 (17):\penalty0 2781--2796, 2021.

\bibitem[Hardcastle et~al.(2015)Hardcastle, Ganguli, and
  Giocomo]{hardcastle2015environmental}
Kiah Hardcastle, Surya Ganguli, and Lisa~M Giocomo.
\newblock Environmental boundaries as an error correction mechanism for grid
  cells.
\newblock \emph{Neuron}, 86\penalty0 (3):\penalty0 827--839, 2015.

\bibitem[Kropff et~al.(2015)Kropff, Carmichael, Moser, and
  Moser]{kropff2015speed}
Emilio Kropff, James~E Carmichael, May-Britt Moser, and Edvard~I Moser.
\newblock Speed cells in the medial entorhinal cortex.
\newblock \emph{Nature}, 523\penalty0 (7561):\penalty0 419--424, 2015.

\bibitem[Langston et~al.(2010)Langston, Ainge, Couey, Canto, Bjerknes, Witter,
  Moser, and Moser]{langston2010development}
Rosamund~F Langston, James~A Ainge, Jonathan~J Couey, Cathrin~B Canto, Tale~L
  Bjerknes, Menno~P Witter, Edvard~I Moser, and May-Britt Moser.
\newblock Development of the spatial representation system in the rat.
\newblock \emph{Science}, 328\penalty0 (5985):\penalty0 1576--1580, 2010.

\bibitem[Dordek et~al.(2016)Dordek, Soudry, Meir, and
  Derdikman]{dordek2016extracting}
Yedidyah Dordek, Daniel Soudry, Ron Meir, and Dori Derdikman.
\newblock Extracting grid cell characteristics from place cell inputs using
  non-negative principal component analysis.
\newblock \emph{Elife}, 5:\penalty0 e10094, 2016.

\bibitem[Barry et~al.(2007)Barry, Hayman, Burgess, and
  Jeffery]{barry2007experience}
Caswell Barry, Robin Hayman, Neil Burgess, and Kathryn~J Jeffery.
\newblock Experience-dependent rescaling of entorhinal grids.
\newblock \emph{Nature neuroscience}, 10\penalty0 (6):\penalty0 682--684, 2007.

\bibitem[Bush et~al.(2014)Bush, Barry, and Burgess]{bush2014grid}
Daniel Bush, Caswell Barry, and Neil Burgess.
\newblock What do grid cells contribute to place cell firing?
\newblock \emph{Trends in neurosciences}, 37\penalty0 (3):\penalty0 136--145,
  2014.

\bibitem[Bao et~al.(2019)Bao, Gjorgieva, Shanahan, Howard, Kahnt, and
  Gottfried]{bao2019grid}
Xiaojun Bao, Eva Gjorgieva, Laura~K Shanahan, James~D Howard, Thorsten Kahnt,
  and Jay~A Gottfried.
\newblock Grid-like neural representations support olfactory navigation of a
  two-dimensional odor space.
\newblock \emph{Neuron}, 102\penalty0 (5):\penalty0 1066--1075, 2019.

\bibitem[Aronov et~al.(2017)Aronov, Nevers, and Tank]{aronov2017mapping}
Dmitriy Aronov, Rhino Nevers, and David~W Tank.
\newblock Mapping of a non-spatial dimension by the hippocampal--entorhinal
  circuit.
\newblock \emph{Nature}, 543\penalty0 (7647):\penalty0 719--722, 2017.

\bibitem[Chen et~al.(2022)Chen, Zhang, Long, and Zhang]{chen2022grid}
Zhe~Sage Chen, Xiaohan Zhang, Xiaoyang Long, and Sheng-Jia Zhang.
\newblock Are grid-like representations a component of all perception and
  cognition?
\newblock \emph{Frontiers in Neural Circuits}, 16, 2022.

\bibitem[Killian et~al.(2012)Killian, Jutras, and Buffalo]{killian2012map}
Nathaniel~J Killian, Michael~J Jutras, and Elizabeth~A Buffalo.
\newblock A map of visual space in the primate entorhinal cortex.
\newblock \emph{Nature}, 491\penalty0 (7426):\penalty0 761--764, 2012.

\bibitem[Julian et~al.(2017)Julian, Keinath, Frazzetta, and
  Epstein]{julian2017evidence}
Joshua Julian, Alex Keinath, Giulia Frazzetta, and Russell Epstein.
\newblock Evidence for a grid-like representation of visual space in humans.
\newblock \emph{Journal of Vision}, 17\penalty0 (10):\penalty0 307--307, 2017.

\bibitem[Wilming et~al.(2018)Wilming, K{\"o}nig, K{\"o}nig, and
  Buffalo]{wilming2018entorhinal}
Niklas Wilming, Peter K{\"o}nig, Seth K{\"o}nig, and Elizabeth~A Buffalo.
\newblock Entorhinal cortex receptive fields are modulated by spatial
  attention, even without movement.
\newblock \emph{Elife}, 7:\penalty0 e31745, 2018.

\bibitem[Bellmund et~al.(2016)Bellmund, Deuker, Schr{\"o}der, and
  Doeller]{bellmund2016grid}
Jacob~LS Bellmund, Lorena Deuker, Tobias~Navarro Schr{\"o}der, and Christian~F
  Doeller.
\newblock Grid-cell representations in mental simulation.
\newblock \emph{Elife}, 5:\penalty0 e17089, 2016.

\bibitem[Constantinescu et~al.(2016)Constantinescu, O’Reilly, and
  Behrens]{constantinescu2016organizing}
Alexandra~O Constantinescu, Jill~X O’Reilly, and Timothy~EJ Behrens.
\newblock Organizing conceptual knowledge in humans with a gridlike code.
\newblock \emph{Science}, 352\penalty0 (6292):\penalty0 1464--1468, 2016.

\bibitem[Vigan{\`o} et~al.(2021)Vigan{\`o}, Rubino, Di~Soccio, Buiatti, and
  Piazza]{vigano2021grid}
Simone Vigan{\`o}, Valerio Rubino, Antonio Di~Soccio, Marco Buiatti, and
  Manuela Piazza.
\newblock Grid-like and distance codes for representing word meaning in the
  human brain.
\newblock \emph{NeuroImage}, 232:\penalty0 117876, 2021.

\bibitem[Eichenbaum(2014)]{eichenbaum2014time}
Howard Eichenbaum.
\newblock Time cells in the hippocampus: a new dimension for mapping memories.
\newblock \emph{Nature Reviews Neuroscience}, 15\penalty0 (11):\penalty0
  732--744, 2014.

\bibitem[Ginosar et~al.(2021)Ginosar, Aljadeff, Burak, Sompolinsky, Las, and
  Ulanovsky]{ginosar2021locally}
Gily Ginosar, Johnatan Aljadeff, Yoram Burak, Haim Sompolinsky, Liora Las, and
  Nachum Ulanovsky.
\newblock Locally ordered representation of 3d space in the entorhinal cortex.
\newblock \emph{Nature}, 596\penalty0 (7872):\penalty0 404--409, 2021.

\bibitem[Grieves et~al.(2021)Grieves, Jedidi-Ayoub, Mishchanchuk, Liu,
  Renaudineau, Duvelle, and Jeffery]{grieves2021irregular}
Roddy~M Grieves, Selim Jedidi-Ayoub, Karyna Mishchanchuk, Anyi Liu, Sophie
  Renaudineau, {\'E}l{\'e}onore Duvelle, and Kate~J Jeffery.
\newblock Irregular distribution of grid cell firing fields in rats exploring a
  3d volumetric space.
\newblock \emph{Nature neuroscience}, 24\penalty0 (11):\penalty0 1567--1573,
  2021.

\bibitem[Hawkins et~al.(2019)Hawkins, Lewis, Klukas, Purdy, and
  Ahmad]{hawkins2019framework}
Jeff Hawkins, Marcus Lewis, Mirko Klukas, Scott Purdy, and Subutai Ahmad.
\newblock A framework for intelligence and cortical function based on grid
  cells in the neocortex.
\newblock \emph{Frontiers in neural circuits}, page 121, 2019.

\bibitem[Bennett et~al.(2009)Bennett, Miller, and Wolford]{bennett2009neural}
Craig~M Bennett, Michael~B Miller, and George~L Wolford.
\newblock Neural correlates of interspecies perspective taking in the
  post-mortem atlantic salmon: An argument for multiple comparisons correction.
\newblock \emph{Neuroimage}, 47\penalty0 (Suppl 1):\penalty0 S125, 2009.

\bibitem[Banino et~al.(2018)Banino, Barry, Uria, Blundell, Lillicrap, Mirowski,
  Pritzel, Chadwick, Degris, Modayil, et~al.]{banino2018vector}
Andrea Banino, Caswell Barry, Benigno Uria, Charles Blundell, Timothy
  Lillicrap, Piotr Mirowski, Alexander Pritzel, Martin~J Chadwick, Thomas
  Degris, Joseph Modayil, et~al.
\newblock Vector-based navigation using grid-like representations in artificial
  agents.
\newblock \emph{Nature}, 557\penalty0 (7705):\penalty0 429--433, 2018.

\bibitem[Banino et~al.(2020)Banino, Kumaran, Hadsell, and
  Benigno]{banino2020performing}
Andrea Banino, Sudarshan Kumaran, Raia~Thais Hadsell, and URIA-MARTINEZ
  Benigno.
\newblock Performing navigation tasks using grid codes, March~31 2020.
\newblock US Patent 10,605,608.

\bibitem[Sorscher et~al.(2019)Sorscher, Mel, Ganguli, and
  Ocko]{sorscher2019unified}
Ben Sorscher, Gabriel Mel, Surya Ganguli, and Samuel Ocko.
\newblock A unified theory for the origin of grid cells through the lens of
  pattern formation.
\newblock \emph{Advances in neural information processing systems}, 32, 2019.

\bibitem[Schaeffer et~al.(2022)Schaeffer, Khona, and Fiete]{schaeffer2022no}
Rylan Schaeffer, Mikail Khona, and Ila Fiete.
\newblock No free lunch from deep learning in neuroscience: A case study
  through models of the entorhinal-hippocampal circuit.
\newblock \emph{bioRxiv}, 2022.

\bibitem[Leadholm et~al.(2021)Leadholm, Lewis, and Ahmad]{leadholm2021grid}
Niels Leadholm, Marcus Lewis, and Subutai Ahmad.
\newblock Grid cell path integration for movement-based visual object
  recognition.
\newblock \emph{arXiv preprint arXiv:2102.09076}, 2021.

\bibitem[Whittington et~al.(2021)Whittington, Warren, and
  Behrens]{whittington2021relating}
James~CR Whittington, Joseph Warren, and Timothy~EJ Behrens.
\newblock Relating transformers to models and neural representations of the
  hippocampal formation.
\newblock \emph{arXiv preprint arXiv:2112.04035}, 2021.

\bibitem[Jonas and Kording(2017)]{jonas2017could}
Eric Jonas and Konrad~Paul Kording.
\newblock Could a neuroscientist understand a microprocessor?
\newblock \emph{PLoS computational biology}, 13\penalty0 (1):\penalty0
  e1005268, 2017.

\bibitem[Hassabis et~al.(2017)Hassabis, Kumaran, Summerfield, and
  Botvinick]{hassabis2017neuroscience}
Demis Hassabis, Dharshan Kumaran, Christopher Summerfield, and Matthew
  Botvinick.
\newblock Neuroscience-inspired artificial intelligence.
\newblock \emph{Neuron}, 95\penalty0 (2):\penalty0 245--258, 2017.

\end{thebibliography}





\end{document}